\documentclass[prb,amsmath,amssymb,superscriptaddress,twocolumn]{revtex4}
\usepackage{graphicx}
\usepackage{bbm,color,ulem}
\DeclareMathAlphabet{\mathpzc}{OT1}{pzc}{m}{it} 
\begin{document}
\title{Spontaneous symmetry breaking in a honeycomb lattice subject to a periodic potential}
\author{Robert E.\ Throckmorton}
\author{S.\ Das Sarma}
\affiliation{Condensed Matter Theory Center and Joint Quantum Institute, Department of Physics, University of Maryland, College Park, Maryland 20742-4111 USA}
\date{\today}
\begin{abstract}
Motivated by recent developments in twisted bilayer graphene moir\'e superlattices, we investigate the effects
of electron-electron interactions in a honeycomb lattice with an applied periodic potential using a finite-temperature
Wilson-Fisher momentum shell renormalization group (RG) approach.  We start with a low-energy effective theory
for such a system, at first giving a discussion of the most general case in which no point group symmetry
is preserved by the applied potential, and then focusing on the special case in which the potential preserves a
$D_3$ point group symmetry.  As in similar studies of bilayer graphene, we find that, while the coupling constants
describing the interactions diverge at or below a certain ``critical temperature'' $T=T_c$, it turns out that {\it ratios}
of these constants remain finite and in fact provide information about what types of orders the system is becoming
unstable to.  However, in contrast to these previous studies, we only find isolated fixed rays, indicating that these
orders are likely unstable to perturbations to the coupling constants.  Our RG analysis leads to the qualitative
conclusion that the emergent interaction-induced symmetry-breaking phases in this model system, and perhaps therefore
by extension in twisted bilayer graphene, are generically unstable and fragile, and may thus manifest strong sample
dependence.
\end{abstract}
\maketitle

\section{Introduction}
Moir\'e superlattices of various kinds have recently become a topic of great theoretical and experimental interest,
spurred in large part by the discovery of superconductivity at a surprisingly high transition temperature\cite{CaoNature2018_1,YankowitzScience2019}
in twisted bilayer graphene (tBLG) with the twist angle close to some small ``magic angle''.  In addition, other
experiments\cite{CaoNature2018_2,BurgPRL2019} have uncovered a Mott-like correlated insulator phase in tBLG, and
various experimental and theoretical investigations of this and other, similar, systems have been undertaken.  For
example, a gate-tunable correlated insulator phase was also found in trilayer graphene moir\'e superlattices\cite{ChenNatPhys2019},
and a crystal field-induced gap in encapsulated twisted double bilayer graphene has been observed\cite{RickhausNanoLett2019}.
Numerous other experiments on tBLG and related systems have also been performed\cite{YankowitzScience2019,JiangArXiv,KerelskyNature2019,JiangNature2019,ChoiNatPhys2019,CaoArXiv,LiuArXiv,ShenArXiv,KerelskyArXiv,WuPRL2018_2,WuPRL2019,WangArXiv,XianNanoLett2019,KariyadoPRR2019,JinNature2019,AndersenArXiv,KennesNatComms2020}.  One of these in
particular\cite{JiangArXiv} is performed on graphene in the presence of a periodic potential induced by buckling of
the sheet, though this also includes an effective periodic magnetic field, which we will not consider in our work.
A consensus that electron-electron interaction, enhanced strongly by the flatband nature of the moir\'e system, is
the driving mechanism producing various symmetry-breaking phases in twisted bilayer graphene has emerged although the
possibility that superconductivity itself may arise from electron-phonon interactions cannot be ruled out\cite{WuPRL2018,WuPRB2019,WuPRB2019_2,PeltonenPRB2018}.
One work proposes a number of potential correlated insulating phases in tBLG\cite{PoPRX2018}.  Another, also considering
tBLG, considers the interplay between van Hove singularities and the symmetry-breaking phases, and proposes different types
of $s$-wave superconductivity\cite{SherunkovPRB2018}.  Yet another\cite{KangPRL2019} considers tBLG at two different fillings,
$1$ and $2$ electrons per moir\'e unit cell, finding a ferromagnetic stripe phase in the former case and an approximately
$SU(4)$ symmetric insulating state in the latter.  Two other works consider other moir\'e superlattice systems, with
one\cite{ZhangPRB2019} describing the existence of Chern bands in twisted double bi- and trilayer graphene and in hexagonal
boron nitride in the presence of an applied electric field, while another\cite{FuArXiv} investigates a number of models
of generic moir\'e systems.  The subject is exhaustive with hundreds of theoretical papers proposing different mechanisms
for symmetry-breaking ground states using many different approximations and models, and we do not attempt a review of the
subject, only mentioning a few representative recent publications where renormalization group (RG) techniques were used to
study the moir\'e ground states\cite{YuanPRB2018,KennesPRB2018,IsobePRX2018,TangPRB2019,LinPRB2019,ChouPRB2019,DasSarmaArXiv}.

We will theoretically investigate, using extensive RG techniques, a related system to those described above---a honeycomb
lattice, such as that formed by graphene, subject to a periodic potential.  While we expect differences with tBLG, we
expect that similar physics should arise generically in both systems.  Our goal is to obtain universal effects of interactions
in the most general situation, necessitating leaving out many important, but nonessential, realistic complications in the
experimental moir\'e systems such as strain, substrates, disorder, band structure details. Such a generic theory using the
minimal model of our work could be the starting point for more realistic future theories.  We therefore restrict ourselves
to the seemingly simple model of a honeycomb lattice subjected to an external periodic potential to mimic the essential
features of twisted bilayer graphene.  As we would see, even this simple system is extremely difficult to handle from the
perspective of dealing with electron-electron interactions because of the greatly reduced symmetries in the interacting
lattice Hamiltonian.

We begin with a low-energy effective theory, which consists of two Dirac cones.  In general, the periodic potential could
completely eliminate all point group symmetry from the system, leaving only (reduced) translational, time reversal, and
spin $SU(2)$ symmetries.  In such a case, we may see terms that shift the Dirac cones away from their usual positions at
the corners of the Brillouin zone in addition to a mass term that opens a gap (there are experimental indications of such
Dirac point gaps in tBLG samples).  Our main focus, however, will be on potentials that preserve a $D_3$ point group symmetry,
in which case only the mass term may appear without any shifts in the Dirac cones.  We then construct the four-fermion
interaction terms that the symmetries of the system allow; we find that $22$ such terms are allowed purely by symmetry, and
that this number may be reduced to $10$ by the use of Fierz identities.  The interaction problem facing us is therefore
formidable involving in general a $22$-parameter RG flow even within this minimal model of this moire system.  Even the reduced
problem of a $10$-parameter flow has never before attempted in the graphene literature.

We employ a finite-temperature Wilson-Fisher momentum shell RG procedure in this work\cite{PolchinskiArXiv,ShankarRMP1994,MillisPRB1993}.
In such a procedure, we rewrite the partition function for our system as a path integral in terms of anticommuting
Grassmann fields and impose a momentum cutoff $\Lambda$ on the electronic modes.  Next, we divide these modes into
``fast'' modes, which are those within a thin shell near the momentum cutoff, and ``slow'' modes, which are the
remaining modes.  We then integrate out the ``fast'' modes perturbatively to one-loop order and rescale the ``slow''
modes and the various coupling constants to recover an action of the same form as that which we started with.  This
procedure yields corrections to the various terms in the action, which yield a set of differential equations that
we call RG equations describing how the various coupling constants evolve as we integrate out modes.  We show that,
if we set the temperature to a ``critical temperature'' $T=T_c$, the constants describing the four-fermion interactions
diverge exponentially, but that {\it ratios} thereof tend to finite values.  Therefore, we find that the coupling
constants tend toward ``fixed rays'' rather than fixed points.  These ratios in fact contain information about what
symmetry-breaking phases the system is unstable to.  To determine which phases these are, we note that, as we integrate
out electronic modes, we also generate contributions to the free energy, which is simply given by $F=-k_BT\ln{Z}$,
where $Z$ is the partition function.  We use this fact to calculate the free energy in the presence of ``source terms''
corresponding to various symmetry-breaking orders; we may then calculate susceptibilities towards these orders by
taking second derivatives with respect to the source term coefficients.  We find that, just above the critical temperature,
any divergent susceptibility does so as a power of $T-T_c$, with the power related to the coupling constant ratios.
If a given susceptibility diverges, then we say that the system is unstable to the associated symmetry-breaking order.
Our work is a highly nontrivial (because of the considerably reduced symmetry of the system leading to multi-parameter
RG flow) generalization of the earlier RG work on bilayer graphene\cite{CvetkovicPRB2012,ThrockmortonPRB2014}. The
imposition of the additional periodic potential considerably complicates the technical aspects of the RG flow compared
with these earlier works, and leads to some qualitative differences in the results as discussed below.

We find that, in contrast to similar studies of bilayer graphene\cite{CvetkovicPRB2012,ThrockmortonPRB2014}, the
only fixed rays that appear here are isolated.  A number of these fixed rays correspond to multiple instabilities.
As a result of these two facts, we expect that the results of integrating the RG equations will be sensitive, even
qualitatively, to changes in initial conditions.  This is not surprising, given the diverse orders found in the
literature in tBLG.  A necessary conclusion is that the symmetry-breaking phases in the system would be fragile and
highly sensitive to all the details of the specific sample, and there could be considerable sample-to-sample variations
in the observed phase diagram (or even in the same sample under thermal cycling).  We emphasize that this finding
of ``unstable symmetry breaking'' in our minimal model , arising from just pristine interaction effects, could only
be much more complex in real tBLG samples where many nonessential realistic effects (e.g. strain, phonon, disorder,
substrate) would come into play well beyond our effective low-energy RG theory.

The rest of this work is organized as follows.  In Sec.\ II, we introduce the system and our low-energy effective
theory.  We then describe and implement our RG procedure in Sec.\ III.  In Sec.\ IV, we describe how we obtain
the fixed rays, and then describe how we determine what instabilities they correspond to in Sec.\ V.  We give
our conclusions in Sec.\ VI, and provide further technical details of the calculation in the appendices.

\section{System and model}
We consider here electrons on a tight-binding honeycomb lattice subject to a periodic potential.  Such
a lattice, in the absence of the periodic potential, possesses a $D_6$ point group symmetry, along with
time reversal, spin $SU(2)$, and discrete translational symmetries.  In the ``worst-case'' scenario,
the potential removes all point group symmetries of the honeycomb lattice, leaving only time reversal,
spin $SU(2)$, and translation symmetries (the last of these reduced by the applied potential).  We assume
that the applied potential is commensurate with the honeycomb lattice.  If the potential is arranged in
such a way as to place a maximum or minimum at a lattice site, however, then the full system will possess
a $D_3$ point group symmetry; we will in fact focus on this case after a brief discussion of the case in
which we have no point group symmetry at all.

We will adopt a low-energy effective theory of this system, which consists of two Dirac cones (valleys
$\pm\vec{K}$), a sublattice degree of freedom (A/B), and spin ($\uparrow$/$\downarrow$).  We also include
all ``mass'' terms allowed by the symmetries of the system.  We list the valley and sublattice components
of all bilinears that may be formed, along with what representations they transform under both with and without
the $D_6$ point group symmetry, in Table \ref{Tab:Classification}.  Some of the bilinears transform nontrivially
under translations; these instead transform under representations of the $D_3$ ``small group'' of the wave
vector, $\vec{K}$.  For the sake of a self-contained presentation, we also list the representations\cite{TinkhamBook}
of $D_6$ and $D_3$ and their associated characters in Tables \ref{Tab:RepsD6} and \ref{Tab:RepsD3}, respectively.
Here, each matrix is of the form $\tau_i\sigma_js_k$, where the $\tau$ matrix operates on the valley pseudospin,
the $\sigma$ matrix on the sublattice pseudospin, and the $s$ matrix on the actual spin.  We should note here
that the form of the low-energy theory is completely independent of the exact details of the applied potential,
though determining the values of the constants appearing therein requires a more detailed calculation.

\begin{table*}
	\centering
		\begin{tabular}{| c | c | c | c |}
			\hline
			\textbf{Matrix $M$} & \textbf{Rep.\ ($D_6$ point group)} & \textbf{Rep.\ ($D_3$ point group)} & \textbf{Rep.\ (no point group)} \\
			\hline\hline
			$1_4$ & $A_1+$ & $A_1+$ & $A+$ \\
			\hline
			$1\sigma_z$ & $B_2+$ & $A_1+$ & $A+$ \\
			\hline
			$\tau_z\sigma_z$ & $A_2-$ & $A_2-$ & $A-$ \\
			\hline
			$\tau_z1$ & $B_1-$ & $A_2-$ & $A-$ \\
			\hline
			$(1\sigma_y,-\tau_z\sigma_x)$ & $E_1-$ & $E-$ & $(A-,A-)$ \\
			\hline
			$(1\sigma_x,\tau_z\sigma_y)$ & $E_2+$ & $E+$ & $(A+,A+)$ \\
			\hline
			$[\tau_x\sigma_x;-\tau_y\sigma_x]$ & $A_{1K}+$ & $A_{1K}+$ & $A_K+$ \\
			\hline
			$[\tau_x\sigma_y;-\tau_y\sigma_y]$ & $A_{2K}-$ & $A_{2K}-$ & $A_K-$ \\
			\hline
			$([\tau_x1;\tau_y1],[\tau_y\sigma_z;-\tau_x\sigma_z])$ & $E_K+$ & $E_K+$ & $(A_K+,A_K+)$ \\
			\hline
		\end{tabular}
		\caption{Bilinears of the form, $\psi^\dag(\vec{r})M1\psi(\vec{r})$, classified according to
		representations of the symmetry group of the honeycomb lattice ($D_6$ point group, spin $SU(2)$,
		time reversal, and translations), those of the reduced symmetry group with only $D_3$ point
		group symmetries, and those with no point group symmetries.  The $+$ or $-$ in each representation
		name indicates how the bilinears transform under time reversal (even or odd, respectively).
		Corresponding to each of these ``charge'' representations are ``spin'' representations, in which
		the spin part of the matrix is $s_k$, $k=x,y,z$; these transform oppositely with respect to
		time reversal to the ``charge'' representations (e.g., the $A_1$ spin representation is odd
		under time reversal).}\label{Tab:Classification}
\end{table*}

\begin{table}
	\centering
		\begin{tabular}{| c | c | c | c | c | c | c |}
			\hline
			\textbf{Representation} & $E$ & $C_2$ & $2C_3$ & $2C_6$ & $3C'_2$ & $3C''_2$ \\
			\hline\hline
			$A_1+$ & $1$ & $1$ & $1$ & $1$ & $1$ & $1$ \\
			\hline
			$A_2-$ & $1$ & $1$ & $1$ & $1$ & $-1$ & $-1$ \\
			\hline
			$B_1-$ & $1$ & $-1$ & $1$ & $-1$ & $1$ & $-1$ \\
			\hline
			$B_2+$ & $1$ & $-1$ & $1$ & $-1$ & $-1$ & $1$ \\
			\hline
			$E_1-$ & $2$ & $-2$ & $-1$ & $1$ & $0$ & $0$ \\
			\hline
			$E_2+$ & $2$ & $2$ & $-1$ & $-1$ & $0$ & $0$ \\
			\hline
		\end{tabular}
		\caption{Representations of the group $D_6$, along with how they transform under time reversal (A ``$+$''
		means the representation is even, while ``$-$'' means that it is odd).}\label{Tab:RepsD6}
\end{table}

\begin{table}
	\centering
		\begin{tabular}{| c | c | c | c |}
			\hline
			\textbf{Representation} & $E$ & $2C_3$ & $C'_2$ \\
			\hline\hline
			$A_1+$ & $1$ & $1$ & $1$ \\
			\hline
			$A_2-$ & $1$ & $1$ & $-1$ \\
			\hline
			$E+$ & $2$ & $-1$ & $0$ \\
			\hline
		\end{tabular}
		\caption{Representations of the group $D_3$, along with how they transform under time reversal (A ``$+$''
		means the representation is even, while ``$-$'' means that it is odd).}\label{Tab:RepsD3}
\end{table}

The model Hamiltonian for the case with no point group symmetry is similar in form to the low-energy effective
theory for a honeycomb lattice, but includes additional terms that transform trivially, i.e., under the $A+$
representation, under the remaining, reduced, symmetry group of the system with an applied periodic potential.
This Hamiltonian is
\begin{eqnarray}
H&=&\sum_{\vec{r}}\psi^\dag(\vec{r})(v_{Fx}p_x\tau_z\sigma_x1-v_{Fy}p_y1\sigma_y1)\psi(\vec{r}) \cr
&+&\sum_{\vec{r}}\psi^\dag(\vec{r})(v_{Fx}k_{0x}1\sigma_x1+v_{Fy}k_{0y}\tau_z\sigma_y1)\psi(\vec{r}) \cr
&+&m\sum_{\vec{r}}\psi^\dag(\vec{r})1\sigma_z1\psi(\vec{r})-\mu\sum_{\vec{r}}\psi^\dag(\vec{r})\psi(\vec{r}).
\end{eqnarray}
The first two ``mass'' terms, the $1\sigma_x1$ and $\tau_z\sigma_y1$ terms, simply displace the Dirac cones
away from $\pm\vec{K}$, while the third, the $1\sigma_z1$ term, opens a gap in the cones.  We discuss the
connection between this low-energy effective Hamiltonian and the exact Hamiltonian that this represents in
Appendix \ref{App:ExactH}.

We have so far covered the noninteracting Hamiltonian; we now turn our attention to the interaction terms.
All such terms must, as with the noninteracting terms, be invariant with respect to the symmetries of the
system.  We may form such invariant terms by use of the generalized Uns\"old theorem; all of these will have
the form,
\begin{equation}
\tfrac{1}{2}\sum_i\sum_{\vec{r}}g_i[\psi^\dag(\vec{r})S_{i,1}\psi(\vec{r})][\psi^\dag(\vec{r})S_{i,2}\psi(\vec{r})],
\end{equation}
where the matrices $S_{i,j}$ both belong to the same ``row'' of a given representation\cite{TinkhamBook}.
In the $D_6$ symmetric case, there is only one matrix per ``row'' for each representation; for example,
\begin{equation}
\tfrac{1}{2}\sum_{\vec{r}}g_{E_2}\{[\psi^\dag(\vec{r})1\sigma_x1\psi(\vec{r})]^2+[\psi^\dag(\vec{r})\tau_z\sigma_y1\psi(\vec{r})]^2\}
\end{equation}
is the only interaction term that we can associate with the $E_2+$ representation.  On the other hand, in the
case of no point group symmetry, we see that there are three matrices in the same ``row'' of, say, the $A+$
representation.  There are therefore six distinct ways to form interaction terms within this representation.
Overall, in the case with no point group symmetry, we find that there are $54$ allowed interaction terms, while,
in the case of $D_3$ point group symmetry, the number is reduced to $22$.  These numbers may be reduced further
through the use of Fierz identities, which are, for $8\times 8$ matrices, of the form,
\begin{eqnarray}
&&(\psi^\dag S_{k,1}\psi)(\psi^\dag S_{k,2}\psi) \cr
&&=-\tfrac{1}{64}\sum_{ij}\mbox{Tr}(S_{k,1}\Lambda_jS_{k,2}\Lambda_i)(\psi^\dag\Lambda_i\psi)(\psi^\dag\Lambda_j\psi),
\end{eqnarray}
where the $\Lambda_n$ are all possible matrices of the form, $\tau_i\sigma_js_k$, and we omitted the position
dependence of the operators $\psi$ for the sake of brevity (we assume that all are at the same position).  As
may be seen, these identities relate the various interaction terms to one another.  Using these identities, we
may reduce the number of independent couplings to $22$ in the case of no point group symmetry and $10$ in the
case of $D_3$ point group symmetry.  Even though we will not consider the $D_6$ case here, we note that such a
symmetry allows $18$ interaction terms, which may be reduced to $9$ using Fierz identities.

\section{Renormalization group (RG) procedure}
We now turn our attention to describing the Wilson-Fisher momentum shell RG technique that we will employ\cite{ShankarRMP1994}.
This technique is as follows.  We begin by writing down the partition function for our system as a path integral:
\begin{equation}
Z=\int D(\psi,\psi^\ast)\,e^{-S(\psi,\psi^\ast)},
\end{equation}
where the $\psi$ are now Grassmann numbers corresponding to the coherent states of the corresponding operators and
the action $S$ is given by
\begin{equation}
S(\psi,\psi^\ast)=\int_0^\beta d\tau\,\psi^\dag(\vec{r},\tau)\left [\frac{\partial}{\partial\tau}+H(\psi,\psi^\ast)\right ]\psi(\vec{r},\tau),
\end{equation}
where $H$ is the interacting Hamiltonian written in ``normal order'' (i.e., Hermitian conjugates to the left of
non-conjugates) and $\tau$ is an ``imaginary time''.  The next step is to integrate out electronic modes in
momentum shells.  We divide the fields into ``slow'' modes, denoted by $\psi_<$, and ``fast'' modes, denoted by
$\psi_>$, where the fast modes are those modes with momenta within the shell, $\Lambda e^{-\delta\ell}\leq k\leq\Lambda$,
and $\delta\ell$ is a small change in a scale parameter $\ell$ used to characterize how many modes have been
integrated out.  Finally, we rescale all momenta of the remaining, slow, fields to $k'=ke^{\delta\ell}$ to restore
the region of integration over momentum to $k'\leq\Lambda$, and then rescale the fields and constants, thus
recovering the original overall form of the action.  This procedure may be done exactly for the noninteracting
system, but must be done perturbatively once we introduce interactions.  We may express this renormalization of
constants in the form of differential equations (RG equations), as we will see shortly.

We now apply this procedure to the system under consideration and, from this point on, we will specialize to the
case of $D_3$ point group symmetry for the sake of relative simplicity.  In this case, $v_{Fx}=v_{Fy}$ and
$k_{0x}=k_{0y}=0$, as the associated terms no longer transform trivially under the symmetries of the system.
We first determine how the various constants determining the theory rescale at ``tree'' level, i.e., at lowest
order in the perturbative expansion.  Performing a Fourier transform, we find that the noninteracting part of
the action $S_0$ is
\begin{eqnarray}
S_0(\psi,\psi^\ast)&=&\int_{\vec{k}\omega}\psi^\dag(\vec{k},\omega)[-i\omega+v_F(\tau_z\sigma_x1k_x-1\sigma_y1k_y) \cr
&+&m1\sigma_z1-\mu]\psi(\vec{k},\omega),
\end{eqnarray}
where $\int_{\vec{k}\omega}$ is a shorthand for the Matsubara sum (here, the sum is over fermionic Matsubara
frequencies, $\omega_n=(2n+1)\frac{\pi}{\beta}$, where $n$ is an integer) and momentum integral,
\begin{equation}
\int_{\vec{k}\omega}=\frac{1}{\beta}\sum_{\omega}\int_{k\leq\Lambda}\frac{d^2\vec{k}}{(2\pi)^2}.
\end{equation}
The interaction terms, on the other hand, are given by
\begin{equation}
S_{\text{int}}=\tfrac{1}{2}\sum_i\int_{1234}g_i[\psi^\dag(1)S_{i,1}\psi(2)][\psi^\dag(3)S_{i,2}\psi(4)],
\end{equation}
where $\int_{1234}$ is a shorthand for the integrals and sums appearing in this expression along with delta
functions expressing momentum and energy conservation,
\begin{eqnarray}
\int_{1234}&=&\frac{1}{\beta^4}\sum_{\omega_1\ldots\omega_4}\int_{k_1,\ldots k_4\leq\Lambda}\frac{d^2\vec{k}_1}{(2\pi)^2}\cdots\frac{d^2\vec{k}_4}{(2\pi)^2} \cr
&\times&(2\pi)^2\delta^{(2)}(\vec{k}_1-\vec{k}_2+\vec{k}_3-\vec{k}_4) \cr
&\times&\beta\delta(\omega_1-\omega_2+\omega_3-\omega_4), \nonumber \\
\end{eqnarray}
and $\psi(n)=\psi(\vec{k}_n,\omega_n)$.

If we now perform the procedure summarized above, we find that the various constants in our theory rescale
at tree level as follows:
\begin{eqnarray}
m'&=&me^{\delta\ell}, \\
\mu'&=&\mu e^{\delta\ell}, \\
g'_{SU}&=&g_{SU}e^{-\delta\ell}, \\
\beta'&=&\beta e^{-\delta\ell}.
\end{eqnarray}
The last may be rewritten in terms of temperature as $T'=Te^{\delta\ell}$.  These may also be recast as
differential equations; letting $x'=x(\ell+\delta\ell)$ and $x=x(\ell)$, where $x$ is any one of the
above constants, we may easily show that
\begin{eqnarray}
\frac{dm}{d\ell}&=&m, \\
\frac{d\mu}{d\ell}&=&\mu, \\
\frac{dg_i}{d\ell}&=&-g_i, \\
\frac{dT}{d\ell}&=&T.
\end{eqnarray}
We thus see that the mass, chemical potential, and temperature are all relevant parameters under RG, that the
Fermi velocity $v_F$ is marginal, and that the four-fermion couplings $g_i$ are irrelevant, all at tree level.
However, the one-loop corrections to these RG equations can, and in some cases will, change these behaviors.

\begin{figure}[htb]
	\centering
		\includegraphics[width=\columnwidth]{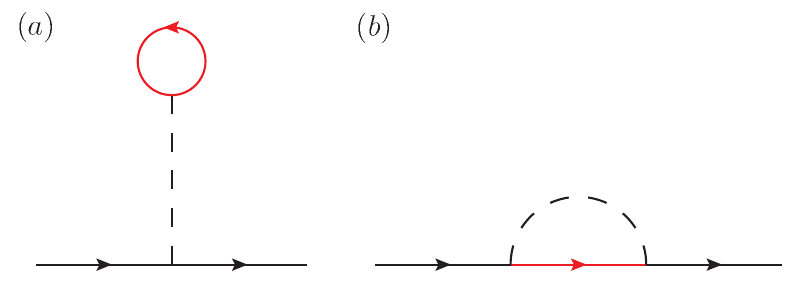}
	\caption{Diagrams representing the one-loop corrections to the chemical potential $\mu$ and mass $m$.  The
	solid red lines represent ``fast'' modes, the solid black lines ``slow'' modes, and the dashed lines interactions.}
	\label{fig:TwoFermion_OneLoop}
\end{figure}

\begin{figure*}[htb]
	\centering
		\includegraphics[width=\textwidth]{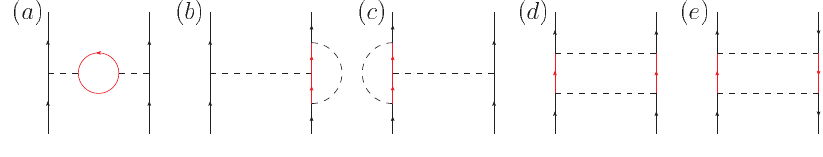}
	\caption{Diagrams representing the one-loop corrections to the four-fermion couplings $g_i$.  The solid red
	lines represent ``fast'' modes, the solid black lines ``slow'' modes, and the dashed lines interactions.}
	\label{fig:FourFermion_OneLoop}
\end{figure*}

We now turn to one-loop corrections.  These are the highest-order corrections that will appear within our
RG analysis, as multi-loop corrections will be of order $(\delta\ell)^k$, $k>1$, and thus will vanish in
the resulting RG equations.  We obtain contributions to $m$, $\mu$, and $g_{SU}$ at this order; we show
the diagrams corresponding to the $m$ and $\mu$ corrections in Fig.\ \ref{fig:TwoFermion_OneLoop} and those
for the corrections to $g_i$ in Fig.\  \ref{fig:FourFermion_OneLoop}.

Before we begin determining these corrections,
we need the bare Green's function for the system $G_0(\vec{k},\omega)$; it is given by
\begin{equation}
G_0(\vec{k},\omega)=\frac{(i\omega+\mu)1_8+v_F(k_x\tau_z\sigma_x1-k_y1\sigma_y1)+m1\sigma_z1}{(-\omega+i\mu)^2+v_F^2k^2+m^2}.
\end{equation}

\subsection{One-loop corrections to mass and chemical potential}
We will begin with the contributions to $m$ and $\mu$, and with the ``tadpole'' diagram, Fig.\ \ref{fig:TwoFermion_OneLoop}a.
Both of these contributions are first order in the interaction terms, and come from those terms containing two ``slow''
and two ``fast'' modes.  This corresponds to terms of the form,
\begin{eqnarray}
&&\Delta S(\text{tadpole}) \cr
&&=\tfrac{1}{2}\sum_ig_i\int_{1234}\langle\psi^\dag_<(1)S_{i,1}\psi_<(2)\psi^\dag_>(3)S_{i,2}\psi_>(4)\rangle_{0,>,C} \cr
&&+(S\longleftrightarrow U).
\end{eqnarray}
Using Wick's theorem and the fact that $G_0(\vec{k},\omega)=\langle\psi(\vec{k},\omega)\psi^\dag(\vec{k},\omega)\rangle$,
we find that
\begin{eqnarray}
&&\Delta S(\text{tadpole}) \cr
&&=-\tfrac{1}{2}\sum_ig_i\int_{1234}\psi^\dag_<(1)S_{i,1}\psi_<(2)\mbox{Tr}[S_{i,2}G(3)]\delta(3-4) \cr
&&+(S_{i,1}\longleftrightarrow S_{i,2}).
\end{eqnarray}
This simplifies to
\begin{eqnarray}
\Delta S(\text{tadpole})&=&-\tfrac{1}{2}\sum_ig_i\int_{\vec{k}\omega}\psi^\dag_<(\vec{k},\omega)S_{i,1}\psi_<(\vec{k},\omega) \cr
&\times&\mbox{Tr}\left [\int_{\vec{k}'\omega'}^>S_{i,2}G(\vec{k}',\omega')\right ]+(S_{i,1}\longleftrightarrow S_{i,2}), \nonumber \\
\end{eqnarray}
where the notation, $\int_{\vec{k}'\omega'}^>$, simply means to integrate only over the ``fast'' momenta.  These sums
and integrals may easily be evaluated; we obtain
\begin{equation}
\int_{\vec{k}\omega}^>\frac{i\omega+\mu}{(\omega+i\mu)^2+[E(\vec{k})]^2}=\frac{\Lambda}{8\pi}F_0(\mu,m,T)\,\delta\ell
\end{equation}
and
\begin{equation}
\int_{\vec{k}\omega}^>\frac{1}{(\omega+i\mu)^2+[E(\vec{k})]^2}=\frac{\Lambda}{8\pi m}F_z(\mu,m,T)\,\delta\ell,
\end{equation}
where
\begin{eqnarray}
F_0(\mu,m;T)&=&\tanh\left (\frac{E_\Lambda-\mu}{2T}\right )-\tanh\left (\frac{E_\Lambda+\mu}{2T}\right ), \nonumber \\ \\
F_z(\mu,m;T)&=&\frac{m}{E_\Lambda}\left [\tanh\left (\frac{E_\Lambda-\mu}{2T}\right )+\tanh\left (\frac{E_\Lambda+\mu}{2T}\right )\right ], \nonumber \\
\end{eqnarray}
and $E_\Lambda=\sqrt{v_F^2\Lambda^2+m^2}$.  If we now substitute this into the expression for $\Delta S(\text{tadpole})$,
we find that
\begin{eqnarray}
&&\Delta S(\text{tadpole}) \cr
&&=-\frac{\Lambda}{16\pi}\sum_ig_i\int_{\vec{k}\omega}\psi^\dag_<(\vec{k},\omega)S_{i,1}\psi_<(\vec{k},\omega) \cr
&&\times[\mbox{Tr}(S_{i,2})F_0(\mu,m;T)+\mbox{Tr}(S_{i,2}1\sigma_z1)F_z(\mu,m;T)]\,\delta\ell \cr
&&+(S\longleftrightarrow U).
\end{eqnarray}
We see that this represents a correction to one of either the chemical potential term or the mass term, depending on whether
$S_{i,1}=1_8$ or $S_{i,1}=1\sigma_z1$.  In fact, the traces in this expression will only be nonzero if $S_{i,2}=1_8$ or
$S_{i,2}=1\sigma_z1$, meaning that only those terms corresponding to the $A_1+$ representation give a nonzero contribution via
the ``tadpole'' diagram.

We now consider the ``sunrise'' diagram, Fig.\ \ref{fig:TwoFermion_OneLoop}b, which corresponds to terms of the form,
\begin{eqnarray}
&&\Delta S(\text{sunrise}) \cr
&&=\frac{1}{2}\sum_ig_i\int_{1234}\langle\psi^\dag_>(1)S_{i,1}\psi_<(2)\psi^\dag_<(3)S_{i,2}\psi_>(4)\rangle_{0,>,C} \cr
&&+\tfrac{1}{2}\sum_ig_i\int_{1234}\langle\psi^\dag_<(1)S_{i,1}\psi_>(2)\psi^\dag_>(3)S_{i,2}\psi_<(4)\rangle_{0,>,C}. \nonumber \\
\end{eqnarray}
Evaluating the averages as before, we get
\begin{eqnarray}
&&\Delta S(\text{sunrise}) \cr
&&=\tfrac{1}{2}\sum_ig_i\int_{\vec{k}\omega}\psi^\dag_<(\vec{k},\omega)\left [\int_{\vec{k}'\omega'}^>S_{i,2}G(\vec{k}',\omega')S_{i,1}\right ]\psi_<(\vec{k},\omega) \cr
&&+(S_{i,1}\longleftrightarrow S_{i,2}).
\end{eqnarray}
Using the formulas given earlier for the integrals over the ``fast'' momenta, this becomes
\begin{eqnarray}
&&\Delta S(\text{sunrise}) \cr
&&=\frac{\Lambda}{16\pi}\sum_ig_i\int_{\vec{k}\omega}\psi^\dag_<(\vec{k},\omega)S_{i,2}S_{i,1}\psi_<(\vec{k},\omega)F_0(\mu,m;T) \cr
&&+\frac{\Lambda}{16\pi}\sum_ig_i\int_{\vec{k}\omega}\psi^\dag_<(\vec{k},\omega)(S_{i,2}1\sigma_z1S_{i,1})\psi_<(\vec{k},\omega)F_z(\mu,m;T) \cr
&&+(S\longleftrightarrow U).
\end{eqnarray}
Unlike the ``tadpole'' contribution, the ``sunrise'' contribution will yield nonzero terms proportional to the coupling constants
coming from \textit{all} representations, not just $A_1+$.

The full RG equations for $\mu$ and $m$ are of the form,
\begin{eqnarray}
\frac{d\mu}{d\ell}&=&\mu+\sum_i[B_{\mu,i}^{0}F_0(\mu,m;T)+B_{\mu,i}^{z}F_z(\mu,m;T)]g_i, \nonumber \\ \\
\frac{dm}{d\ell}&=&m+\sum_i[B_{m,i}^{0}F_0(\mu,m;T)+B_{m,i}^{z}F_z(\mu,m;T)]g_i. \nonumber \\
\end{eqnarray}
When we evaluate the coefficients, however, we find that all of the $B_{\mu,i}^z=0$ and that all of the $B_{m,i}^0=0$.  This simplifies
our later calculations, as we will see below.  This also implies, as may be seen from the form of the equations, that, if we set either
$\mu$ or $m$ to zero, then we will never generate them, i.e., they will not renormalize to nonzero values.

\subsection{One-loop corrections to four-fermion coupling constants}
We next determine the corrections to the four-fermion coupling constants $g_{SU}$, which are depicted in Fig.\ \ref{fig:FourFermion_OneLoop}.
Evaluating these five contributions, we find that the RG equations for the four-fermion couplings all have the form,
\begin{equation}
\frac{dg_i}{d\ell}=-g_i+\sum_{j,k}\sum_{a}A^{(a)}_{ijk}g_jg_k\Phi_a(\mu,m;T). \label{Eq:gRGEqus}
\end{equation}
In arriving at this form, we evaluate integrals and sums of the form,
\begin{equation}
\int_{\vec{k}\omega}^>G_0(\vec{k},\omega)\otimes G_0(\pm\vec{k},\pm\omega). \label{Eq:2GFIntSum}
\end{equation}
We present the results of doing so, in the form of the functions $\Phi_a$, in Appendix \ref{App:Ints_gRGEqus}.  We list the expressions
obtained from each of the diagrams in Fig.\ \ref{fig:FourFermion_OneLoop} in Appendix \ref{App:Coeffs_gRGEqus}.

\section{Fixed rays}
We now determine what the various outcomes of integrating the RG equations are.  We start by showing that, if the temperature is tuned to
what we will call the critical temperature, $T=T_c$, integrating the RG equations for the four-fermion couplings $g_i$ will result in at least
one of said couplings diverging exponentially.  However, it turns out that {\it ratios} of these couplings tend toward fixed values; we will
show later that these ``fixed rays'' in the space of the $g_i$ tell us what symmetry-breaking orders the system is unstable to.  If the
temperature is above this critical temperature, then the $g_i$ all saturate to finite values as $\ell\to\infty$.  If, on the other hand, the
temperature is below the critical temperature, then one or more of the $g_i$ will diverge to infinity at some finite value of $\ell$.

In all of these cases, we need to first solve for the fixed ratios themselves.  To do this, we first derive the equations for ratios of the
$g_i$ with one of the couplings, which we will call $g_r$ (which of course is assumed to diverge),
\begin{equation}
\rho_i=\frac{g_i}{g_r}.
\end{equation}
Doing this, we obtain
\begin{eqnarray}
\frac{d\rho_i}{d\ell}&=&\frac{1}{g_r^2}\left (\frac{dg_i}{d\ell}g_r-g_i\frac{dg_r}{d\ell}\right ) \cr
&=&g_r\sum_{j,k}\sum_{a}[A^{(a)}_{ijk}-A^{(a)}_{rjk}\rho_i]\rho_j\rho_k\Phi_a(\mu,m;T). \nonumber \\ \label{Eq:RatioEqn}
\end{eqnarray}

We find that the behavior of the equations for large $\ell$ depends on how rapidly $T$, $\mu$, and $m$ increase, and breaks down into three
cases:
\begin{enumerate}
	\item $T$ runs faster than $\mu$ and $m$.
	\item $\mu$ and $m$ run faster than $T$ and $|\mu|>|m|$.
	\item $\mu$ and $m$ run faster than $T$ and $|\mu|<|m|$.
\end{enumerate}
Which of these three cases we consider determines the form of the equations that we have to solve to determine the fixed rays and what instabilities
they represent.  We now discuss each case in turn.

{\bf Case 1}: In this case, the temperature $T$ increases more quickly than $\mu$ or $m$ for large $\ell$.  We will determine the large $\ell$
behavior of $\mu$, $m$, and the $g_i$ under this assumption.  In the limit of large $\ell$, the $\Phi_a$ functions that decrease the most slowly
are
\begin{equation}
\Phi_{2,+}\approx\Phi_{2,-}\approx\frac{\Lambda}{8\pi T}.
\end{equation}
In addition, the functions $F_0$ and $F_z$ appearing in the equations for $\mu$ and $m$ behave as follows.
\begin{eqnarray}
F_0\approx\frac{\Lambda}{8\pi}\frac{\mu}{T}, \\
F_z\approx\frac{\Lambda}{8\pi}\frac{m}{T}.
\end{eqnarray}
We first determine the asymptotic behavior of the four-fermion couplings $g_i$ for $T=T_c$.  The equations will all take the form,
\begin{equation}
\frac{dg_i}{d\ell}=-g_i+\frac{\Lambda e^{-\ell}}{8\pi T_c}\sum_{jk}\bar{A}_{ijk}g_jg_k,
\end{equation}
where $\bar{A}_{ijk}=A^{2,+}_{ijk}+A^{2,-}_{ijk}$.  By the definition of $T_c$, we know that, for $T=T_c$, $g_i(\ell\to\infty)\to\infty$.  We
now substitute in $g_i(\ell)=g_{i,0}e^{\delta_g\ell}$, obtaining
\begin{equation}
\delta_gg_{i,0}e^{\delta_g\ell}=-g_{i,0}e^{\delta_g\ell}+\frac{\Lambda e^{(2\delta_g-1)\ell}}{8\pi T_c}\sum_{jk}\bar{A}_{ijk}g_{j,0}g_{k,0},
\end{equation}
or
\begin{equation}
(\delta_g+1)g_{i,0}e^{\delta_g\ell}=\frac{\Lambda e^{(2\delta_g-1)\ell}}{8\pi T_c}\sum_{jk}\bar{A}_{ijk}g_{j,0}g_{k,0}.
\end{equation}
This equation is satisfied if $\delta_g=2\delta_g-1$, or $\delta_g=1$.  This is consistent with our assertion that $g_i$ diverges to
infinity as $\ell\to\infty$.  We now determine the constants, $g_{i,0}$.  We mentioned earlier that ratios of any two divergent $g_i$
tend to finite values; we will rewrite the above equation in terms of these ratios.  If we now rewrite the above equation in terms of
these ratios $\rho_i$, we obtain, after simplification,
\begin{equation}
2g_{r,0}=\frac{\Lambda}{8\pi T_c}g^2_{r,0}\sum_{jk}\bar{A}_{rjk}\rho_j\rho_k.
\end{equation}
Solving for $g_{r,0}$, we obtain
\begin{equation}
g_{r,0}=\frac{16\pi T_c}{\Lambda\mathcal{A}_r},
\end{equation}
where
\begin{equation}
\mathcal{A}_r=\sum_{jk}\bar{A}_{rjk}\rho_j\rho_k.
\end{equation}
All other $g_{i,0}$ may be obtained simply by multiplying $g_{r,0}$ by the appropriate ratio.

Next, we consider the equations for $\mu$ and $m$.  In the limit of large $\ell$, these become
\begin{eqnarray}
\frac{d\mu}{d\ell}&=&\mu+\frac{\Lambda e^{-\ell}}{8\pi T_c}\sum_i(B_{\mu,i}^{0}\mu+B_{\mu,i}^{z}m)g_i, \\
\frac{dm}{d\ell}&=&m+\frac{\Lambda e^{-\ell}}{8\pi T_c}\sum_i(B_{m,i}^{0}\mu+B_{m,i}^{z}m)g_i.
\end{eqnarray}
If we substitute the asymptotic expressions for $g_i$ obtained above into these equations, we get
\begin{eqnarray}
\frac{d\mu}{d\ell}&=&\mu+\frac{2}{\mathcal{A}_r}(\mathcal{B}_\mu^0\mu+\mathcal{B}_\mu^zm), \\
\frac{dm}{d\ell}&=&m+\frac{2}{\mathcal{A}_r}(\mathcal{B}_m^0\mu+\mathcal{B}_m^zm),
\end{eqnarray}
where
\begin{equation}
\mathcal{B}_x^y=\sum_iB_{x,i}^y\rho_i.
\end{equation}
We see that the equations for $\mu$ and $m$ reduce to a pair of first-order linear differential equations.  Our earlier
results imply that $\mathcal{B}_\mu^z=\mathcal{B}_m^0=0$, so that these equations are decoupled.  We just need one of
either $\mu$ or $m$ to increase more slowly than $T$ (or even decrease); if the other increases more quickly, then that
means that the corresponding parameter must be set to zero to obtain that outcome.

{\bf Case 2}: Next, we consider the case in which the $g_i$ increases more slowly than $\mu$ and $m$, and in which $|\mu|>|m|$, i.e.,
the chemical potential is outside the gap.  In this case, the most slowly-increasing of the $\Phi_a$ functions will be
\begin{eqnarray}
\Phi_{0,z,-}&\approx&\frac{\Lambda}{8\pi}\frac{m^2}{\mu(\mu^2-m^2)}, \\
\Phi_{1,z,-+}&=&-\Phi_{1,z,--}\approx -\frac{\Lambda}{8\pi}\frac{m}{\mu^2-m^2}, \\
\Phi_{2,-}&\approx& -\frac{\Lambda}{8\pi}\frac{2\mu^2-m^2}{\mu(\mu^2-m^2)}.
\end{eqnarray}
We will assume for now that the $g_i$, $\mu$, and $m$ all increase exponentially for large $\ell$; we will show here that this assumption
is consistent.  For concreteness, we assume that $g_i(\ell)=g_{i,0}e^{\delta_g\ell}$, $\mu(\ell)=\mu_0 e^{\eta\ell}$, and $m(\ell)=m_0e^{\eta\ell}$.
In this case, the equations for $g_i$ become
\begin{eqnarray}
\frac{dg_i}{d\ell}&=&-g_i+\frac{\Lambda}{8\pi}\sum_{j,k}\left [A^{(0,z,-)}_{ijk}\frac{m_0^2}{\mu_0(\mu_0^2-m_0^2)}\right. \cr
&-&(A^{(1,z,-+)}_{ijk}-A^{(1,z,--)}_{ijk})\frac{m_0}{\mu_0^2-m_0^2} \cr
&-&\left.A^{(2,-)}_{ijk}\frac{2\mu_0^2-m_0^2}{\mu_0(\mu_0^2-m_0^2)}\right ]e^{-\eta\ell}g_jg_k.
\end{eqnarray}
Substituting our ansatz for $g_i$ into this equation, we get
\begin{eqnarray}
\delta_gg_{i,0}e^{\delta_g\ell}&=&-g_{i,0}e^{\delta_g\ell}+\frac{\Lambda}{8\pi}\sum_{j,k}\left [A^{(0,z,-)}_{ijk}\frac{m_0^2}{\mu_0(\mu_0^2-m_0^2)}\right. \cr
&-&(A^{(1,z,-+)}_{ijk}-A^{(1,z,--)}_{ijk})\frac{m_0}{\mu_0^2-m_0^2} \cr
&-&\left.A^{(2,-)}_{ijk}\frac{2\mu_0^2-m_0^2}{\mu_0(\mu_0^2-m_0^2)}\right ]g_{j,0}g_{k,0}e^{(2\delta_g-\eta)\ell}.
\end{eqnarray}
This equation is satisfied if we let $\delta_g=2\delta_g-\eta$, or $\delta_g=\eta$.  We may now solve for $g_{i,0}$ with a similar procedure to
the previous case.  Rewriting the above equation in terms of the ratios $\rho_i$ and simplifying, we get
\begin{eqnarray}
\frac{\eta+1}{g_{r,0}}&=&\frac{\Lambda}{8\pi(\mu_0^2-m_0^2)}\sum_{j,k}\left [A^{(0,z,-)}_{rjk}\frac{m_0^2}{\mu_0}\right. \cr
&-&(A^{(1,z,-+)}_{rjk}-A^{(1,z,--)}_{rjk})m_0 \cr
&-&\left.A^{(2,-)}_{rjk}\frac{2\mu_0^2-m_0^2}{\mu_0}\right ]\rho_j\rho_k.
\end{eqnarray}
If we now denote the sum over $j$ and $k$ by $\mathcal{A}'_{r}(\mu_0,m_0)$, we get
\begin{equation}
g_{r,0}=\frac{8\pi(\eta+1)(\mu_0^2-m_0^2)}{\Lambda\mathcal{A}'_{r}(\mu_0,m_0)}.
\end{equation}
As before, we may obtain the other $g_{i,0}$ by multiplying the above by the appropriate ratio.

With this result, we can now consider the equations for $\mu$ and $m$.  In the limit of large $\ell$, only $F_0$ is nonzero:
\begin{equation}
F_0(\mu,m;T)\approx\frac{\Lambda}{4\pi}\mbox{sgn}\left (\frac{|m|+\mu}{2T}\right ).
\end{equation}
We then obtain
\begin{eqnarray}
\frac{d\mu}{d\ell}&=&\mu+\frac{\Lambda}{4\pi}\mbox{sgn}\left (\frac{|m|+\mu}{2T}\right )\sum_iB_{\mu,i}^{0}g_i, \nonumber \\ \\
\frac{dm}{d\ell}&=&m+\frac{\Lambda}{4\pi}\mbox{sgn}\left (\frac{|m|+\mu}{2T}\right )\sum_iB_{m,i}^{0}g_i. \nonumber \\
\end{eqnarray}
If we now make our earlier ansatz and substitute the expression for the $g_i$ obtained earlier, we obtain
\begin{eqnarray}
(\eta-1)\mu_0&=&\frac{2(\eta+1)(\mu_0^2-m_0^2)}{\mathcal{A}'_{r}(\mu_0,m_0)}\mbox{sgn}\left (\frac{|m_0|+\mu_0}{2T}\right ) \cr
&\times&\sum_iB_{\mu,i}^{0}\rho_i, \nonumber \\ \\
(\eta-1)m_0&=&\frac{2(\eta+1)(\mu_0^2-m_0^2)}{\mathcal{A}'_{r}(\mu_0,m_0)}\mbox{sgn}\left (\frac{|m_0|+\mu_0}{2T}\right ) \cr
&\times&\sum_iB_{m,i}^{0}\rho_i. \nonumber \\
\end{eqnarray}
We note, however, that $B_{m,i}^0=0$, so that the second equation simplifies to
\begin{eqnarray}
(\eta-1)\mu_0&=&\frac{2(\eta+1)(\mu_0^2-m_0^2)}{\mathcal{A}'_{r}(\mu_0,m_0)}\mbox{sgn}\left (\frac{|m_0|+\mu_0}{2T}\right ) \cr
&\times&\sum_iB_{\mu,i}^{0}\rho_i, \nonumber \\ \\
(\eta-1)m_0&=&0.
\end{eqnarray}
We thus find that, unless we set $m_0=0$, we must take $\eta=1$, violating our assumption that $T$ increases more
slowly than $\mu$ or $m$.  If we take $m_0=0$, then the equation for $\mu$ becomes
\begin{equation}
(\eta-1)\mu_0=\frac{2(\eta+1)\mu_0^2}{\mathcal{A}'_{r}(\mu_0,0)}\mbox{sgn}\left (\frac{\mu_0}{2T}\right )\sum_iB_{\mu,i}^{0}\rho_i.
\end{equation}
However, our expression for $\mathcal{A}'_{r}(\mu_0,m_0)$ reduces to
\begin{equation}
\mathcal{A}'_{r}(\mu_0,0)=-2\mu_0\sum_{j,k}A^{(2,-)}_{rjk}\rho_j\rho_k.
\end{equation}
We expect the most likely outcomes of actual integration of the RG equations to fall within the previous case,
and thus we will not treat this case further here.  In concluding this, we are guided by a similar study of
bilayer graphene with a band gap opened by, for example, an applied electric field undertaken in Ref.\ \onlinecite{ThrockmortonPRB2014}.
We also note that the requirement that $m_0=0$ would imply a complete absence of a periodic potential, i.e.,
we are dealing with just a honeycomb lattice.

{\bf Case 3}:
Finally, we consider the case in which, once again, $\mu$ and $m$ increase more rapidly than $T$ for large $\ell$, but this time
$|\mu|<|m|$, i.e., the chemical potential is inside the gap.  In this case, the most slowly-decreasing of the $\Phi_a$ are
\begin{eqnarray}
\Phi_{0,z,+}&\approx&\frac{\Lambda}{8\pi|m|}, \\
\Phi_{2,+}&\approx&\frac{\Lambda}{8\pi|m|}, \\
\Phi_{0,z,-}&\approx&\frac{\Lambda}{8\pi}\frac{|m|}{m^2-\mu^2}, \\
\Phi_{1,z,-+}&=&-\Phi_{1,z,--}\approx-\frac{\Lambda}{8\pi}\frac{\mu}{m^2-\mu^2}\mbox{sgn }{m}, \\
\Phi_{2,-}&\approx&-\frac{\Lambda}{8\pi}\frac{|m|}{m^2-\mu^2}.
\end{eqnarray}
If we now make the same ansatz as before, taking $g_i(\ell)=g_{i,0}e^{\delta_g\ell}$, $\mu(\ell)=\mu_0e^{\eta\ell}$, and
$m(\ell)=m_0e^{\eta\ell}$, then the equations for the $g_i$ become
\begin{eqnarray}
(\delta_g+1)g_{i,0}e^{\delta_g\ell}&=&\frac{\Lambda e^{(2\delta_g-\eta)\ell}}{8\pi|m_0|}\sum_{j,k}\left [(A^{(0,z,+)}_{ijk}+A^{(2,+)}_{ijk})\right. \cr
&+&\frac{m_0^2}{m_0^2-\mu_0^2}(A^{(0,z,-)}_{ijk}+A^{(2,-)}_{ijk}) \cr
&-&\left.\frac{\mu_0 m_0}{m_0^2-\mu_0^2}(A^{(1,z,-+)}_{ijk}+A^{(1,z,--)}_{ijk})\right ]g_{j,0}g_{k,0}. \nonumber \\
\end{eqnarray}
Once again, this equation is satisfied if $\delta_g=\eta$.  We may solve for this in terms of the fixed ratios as before, obtaining
\begin{eqnarray}
\frac{\delta_g+1}{g_{r,0}}&=&\frac{\Lambda}{8\pi|m_0|}\sum_{j,k}\left [(A^{(0,z,+)}_{rjk}+A^{(2,+)}_{rjk})\right. \cr
&+&\frac{m_0^2}{m_0^2-\mu_0^2}(A^{(0,z,-)}_{rjk}+A^{(2,-)}_{rjk}) \cr
&-&\left.\frac{\mu_0 m_0}{m_0^2-\mu_0^2}(A^{(1,z,-+)}_{rjk}+A^{(1,z,--)}_{rjk})\right ]\rho_j\rho_k, \nonumber \\
\end{eqnarray}
or, denoting the sum on $j$ and $k$ by $\mathcal{A}''_r(\mu_0,m_0)$,
\begin{equation}
g_{r,0}=\frac{8\pi(\delta_g+1)|m_0|}{\Lambda\mathcal{A}''_r(\mu_0,m_0)}.
\end{equation}

Now we consider the equations for $\mu$ and $m$.  In this case, only $F_z$ is nonzero for large $\ell$:
\begin{equation}
F_z(\mu,m;T)\approx\frac{\Lambda}{4\pi}\mbox{sgn }m.
\end{equation}
The equations then become
\begin{eqnarray}
\frac{d\mu}{d\ell}&=&\mu+\frac{\Lambda}{4\pi}\mbox{sgn }m_0\sum_iB_{\mu,i}^{z}g_i, \nonumber \\ \\
\frac{dm}{d\ell}&=&m+\frac{\Lambda}{4\pi}\mbox{sgn }m_0\sum_iB_{m,i}^{z}g_i. \nonumber \\
\end{eqnarray}
If we now make our earlier ansatz and substitute the expression for the $g_i$ obtained earlier, we obtain
\begin{eqnarray}
(\eta-1)\mu_0&=&\frac{2(\eta+1)|m_0|}{\mathcal{A}''_{r}(\mu_0,m_0)}\mbox{sgn }m_0\sum_iB_{\mu,i}^{z}\rho_i, \nonumber \\ \\
(\eta-1)m_0&=&\frac{2(\eta+1)|m_0|}{\mathcal{A}''_{r}(\mu_0,m_0)}\mbox{sgn }m_0\sum_iB_{m,i}^{z}\rho_i. \nonumber \\
\end{eqnarray}
We note, however, that $B_{\mu,i}^z=0$, and thus the above becomes
\begin{eqnarray}
(\eta-1)\mu_0&=&0, \\
(\eta-1)m_0&=&\frac{2(\eta+1)|m_0|}{\mathcal{A}''_{r}(\mu_0,m_0)}\mbox{sgn }m_0\sum_iB_{m,i}^{z}\rho_i. \nonumber \\
\end{eqnarray}
Similarly to the previous case, we conclude that we must set $\mu_0=0$ in order to obtain consistency with
our assumptions in this case.  if we do so, however, we find that $\mathcal{A}''_{r}(\mu_0,m_0)$ simplifies
to
\begin{eqnarray}
\mathcal{A}''_{r}(0,m_0)&=&\sum_{j,k}\left (A^{(0,z,+)}_{rjk}+A^{(2,+)}_{rjk}\right. \cr
&+&\left.A^{(0,z,-)}_{rjk}+A^{(2,-)}_{rjk}\right )\rho_j\rho_k.
\end{eqnarray}
For similar reasons as in the previous case, we will save further treatment of this case for future work.

\section{Analysis of fixed rays}
We now describe how we determine what symmetry-breaking phases each fixed ray represents an instability towards.
To do this, we calculate the susceptibility of the system as a function of temperature near the ``critical
temperature''.  We can, in turn, do this by noting that, as we integrate out electronic modes in our RG analysis,
we produce a multiplicative constant contribution to the partition function that we have been ignoring so far.
These multiplicative constants in fact represent contributions to the free energy of the system.  Our basic
strategy for determining the susceptibilities is as follows.  We start by adding ``source terms'' to the action,
which have the form,
\begin{eqnarray}
S_\Delta&=&\sum_i\Delta_i^\text{ph}\int_{\vec{k}\omega}\psi^\dag(\vec{k},\omega)M_i\psi(\vec{k},\omega) \cr
&+&\tfrac{1}{2}\sum_i\Delta_i^\text{pp}\int_{\vec{k}\omega}\psi^\dag(\vec{k},\omega)M_i\psi^\ast(\vec{k},\omega)+\text{c.c.},
\end{eqnarray}
where the matrices $M_i$ run over all possible $8\times 8$ matrices of the form, $\tau_i\otimes\sigma_j\otimes s_k$.
Note that some of the $\Delta$ may be equal to one another if their associated matrices belong to the same
representation of $D_3$.  These source terms correspond to different ``particle-hole'' (ph), or excitonic, and
``particle-particle'' (pp), or superconducting, order parameters.  We provide a list of the representations that
each of these source terms correspond to, along with what order they represent, in Tables \ref{tab:PHOrderParameters}
(ph) and \ref{tab:PPOrderParameters} (pp).  We note that the excitonic states are very similar to those that
are possible in bilayer graphene\cite{CvetkovicPRB2012} due to the mathematical similarity to that case, though
the physical interpretation will be slightly different.
\begin{table*}
	\centering
		\begin{tabular}{| c | c | c | c |}
			\hline
			\textbf{Representation} & \textbf{Matrices} & \textbf{Charge order} & \textbf{Spin order} \\
			\hline
			\hline
			$A_1+$ & $1_4,1\sigma_z$ & Charge density wave & Ferrimagnetic \\
			\hline
			$A_2-$ & $\tau_z\sigma_z,\tau_z1$ & Anomalous quantum Hall & Quantum spin Hall \\
			\hline
			$E_1-$ & $(1\sigma_y,-\tau_z\sigma_x)$ & Loop current & Loop spin current \\
			\hline
			$E_2+$ & $(1\sigma_x,\tau_z\sigma_y)$ & Nematic & Spin nematic \\
			\hline
			$A_{1K}+$ & $[\tau_x\sigma_x;-\tau_y\sigma_x]$ & Kekul\'e & Spin Kekul\'e \\
			\hline
			$A_{2K}-$ & $[\tau_x\sigma_y;-\tau_y\sigma_y]$ & Kekul\'e current & Kekul\'e spin current \\
			\hline
			$E_K+$ & $([\tau_x1;\tau_y1],[\tau_y\sigma_z;-\tau_x\sigma_z])$ & Charge density wave & Spin density wave \\
			\hline
		\end{tabular}
	\caption{List of representations of $D_3$ and the corresponding particle-hole (excitonic) order parameters.  The $\pm$
		after each representation represents how the corresponding charge order transforms under time reversal.  Only the	valley
		and sublattice components of the matrices are shown.  We list both the charge and spin variants of each order in the same
		row; the spin order has the opposite time reversal symmetry to the corresponding charge order (e.g., the ferrimagnetic
		state is {\it odd} under time reversal).}
	\label{tab:PHOrderParameters}
\end{table*}
\begin{table*}
	\centering
		\begin{tabular}{| c | c | c |}
			\hline
			\textbf{Representation} & \textbf{Matrices} & \textbf{Order} \\
			\hline
			\hline
			$A_1$ (s) & $\tau_x1,\tau_x\sigma_z$ & $s+d_{z^2-x^2-y^2}$-wave \\
			\hline
			$A_2$ (t) & $\tau_y\sigma_z,\tau_y1$ & $p_z+f$-wave \\
			\hline
			$E_1$ (t) & $(\tau_y\sigma_x,\tau_x\sigma_y)$ & $(p_x,p_y)$-wave \\
			\hline
			$E_2$ (s) & $(\tau_x\sigma_x,\tau_y\sigma_y)$ & $(d_{x^2-y^2},d_{xy})$-wave \\
			\hline
			$A_{1K}$ (s) & $[1\sigma_x,\tau_z\sigma_x]$ & $s$-wave PDW \\
			\hline
			$A_{2K}$ (t) & $[1\sigma_y,\tau_z\sigma_y]$ & $p_z$-wave PDW \\
			\hline
			$E_K$ (s) & $([1_4;1\sigma_z],[\tau_z\sigma_z;-\tau_z1])$ & $(d_{x^2-y^2},d_{xy},d_{xz},d_{yz})$-wave PDW \\
			\hline
		\end{tabular}
	\caption{List of representations of $D_3$ and the corresponding particle-particle (superconducting) order parameters.  The
	letter after each representation name denotes whether the order is a singlet (s) or triplet (t) order.  We omit the spin
	matrix in the list of matrices; it is $s_y$ for singlet orders and $1$, $s_x$, or $s_z$ for triplet orders.}
	\label{tab:PPOrderParameters}
\end{table*}
In the pp case, only those terms with antisymmetric matrices $M_i$ appear.  We determine the contributions to the free
energy as we integrate out modes up to second order in these source terms.  Finally, we can calculate the susceptibilities
to various order parameters by calculating second derivatives of the free energy:
\begin{eqnarray}
\chi_{ij}^\text{ph}&=&-\frac{\partial^2f}{\partial\Delta_i^\text{ph}\partial\Delta_j^\text{ph}}, \\
\chi_{ij}^\text{pp}&=&-\frac{\partial^2f}{\partial(\Delta_i^\text{pp})^\ast\partial\Delta_j^\text{pp}},
\end{eqnarray}
where $f$ is the free energy per unit area.

\subsection{One-loop RG equations for the source terms}
\begin{figure}
	\centering
		\includegraphics[width=\columnwidth]{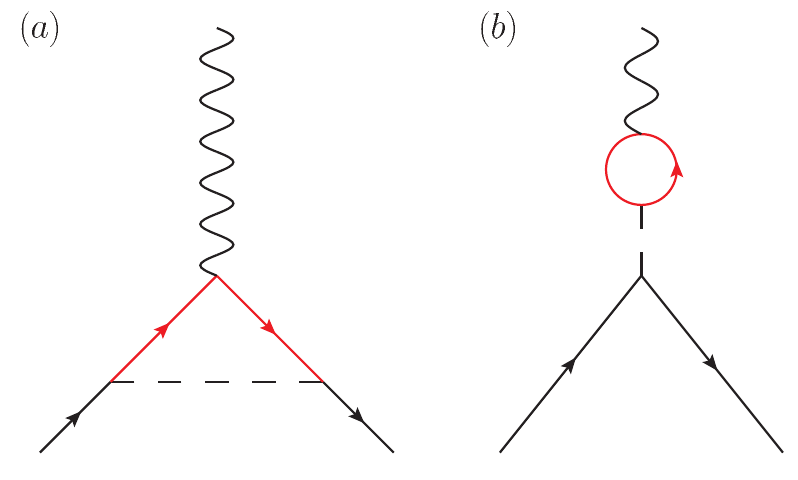}
	\caption{Diagrams corresponding to one-loop corrections to the particle-hole (ph) source terms.  Solid black lines correspond to
	``slow'' modes, solid red lines to ``fast'' modes, dashed lines to four-fermion interactions, and wavy lines to a source term vertex.}
	\label{fig:PHSourceTerms_OneLoop}
\end{figure}

\begin{figure}
	\centering
		\includegraphics[width=0.5\columnwidth]{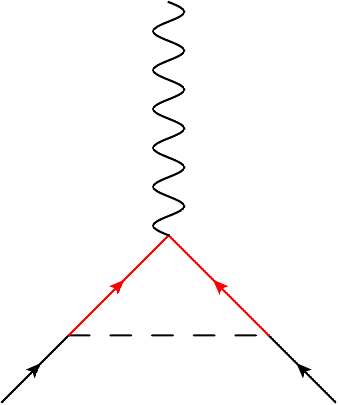}
	\caption{Diagram corresponding to one-loop corrections to the particle-particle (pp) source terms.  Solid black lines correspond to
	``slow'' modes, solid red lines to ``fast'' modes, dashed lines to four-fermion interactions, and wavy lines to a source term vertex.}
	\label{fig:PPSourceTerms_OneLoop}
\end{figure}

Before we determine the free energy, we must also determine how these source terms renormalize to one loop.  At
tree level, the renormalized source terms are given by $\Delta(\ell)=\Delta_0 e^{\ell}$, or
\begin{equation}
\frac{d\Delta}{d\ell}=\Delta.
\end{equation}
We now determine the one-loop corrections, depicted in Figs.\ \ref{fig:PHSourceTerms_OneLoop} and \ref{fig:PPSourceTerms_OneLoop}.
The corrections for the ph source terms are given in Fig.\ \ref{fig:PHSourceTerms_OneLoop}.  The first diagram, Fig.\ \ref{fig:PHSourceTerms_OneLoop}a,
yields
\begin{eqnarray}
\Delta S_\Delta^{\text{ph}}(3a)&=&\tfrac{1}{4}\int_{\vec{k}\omega}\sum_{ij}\Delta_j^{\text{ph}}\psi^\dag(\vec{k},\omega)S_{i,2}\bar{M}_jS_{i,1}\psi(\vec{k},\omega) \cr
&+&(S_{i,1}\longleftrightarrow S_{i,2}),
\end{eqnarray}
where
\begin{equation}
\bar{M}_j=\int_{\vec{k}'\omega'}^>G(\vec{k}',\omega')M_jG(\vec{k}',\omega').
\end{equation}
The second, Fig.\ \ref{fig:PHSourceTerms_OneLoop}b, yields
\begin{eqnarray}
\Delta S_\Delta^{\text{ph}}(3b)&=&-\tfrac{1}{4}\int_{\vec{k}\omega}\sum_{ij}\Delta_j^{\text{ph}}T_{ij}\psi^\dag(\vec{k},\omega)S_{i,1}\psi(\vec{k},\omega) \cr
&+&(S_{i,1}\longleftrightarrow S_{i,2}),
\end{eqnarray}
where
\begin{equation}
T_{ij}=\mbox{Tr}\left [\int_{\vec{k}'\omega'}^>G(\vec{k}',\omega')S_{i,2}G(\vec{k}',\omega')M_j\right ].
\end{equation}
Now we consider the pp source terms.  In this case, we have only one diagram contributing to one-loop renormalization, shown in
Fig.\ \ref{fig:PPSourceTerms_OneLoop}.  This diagram yields
\begin{eqnarray}
\Delta S_\Delta^{\text{pp}}&=&\tfrac{1}{4}\int_{\vec{k}\omega}\sum_{ij}\Delta_j^{\text{ph}}\psi^\dag(\vec{k},\omega)S_{i,1}\bar{N}_jS_{i,2}^T\psi(\vec{k},\omega) \cr
&+&(S_{i,1}\longleftrightarrow S_{i,2}),
\end{eqnarray}
where
\begin{equation}
\bar{N}_j=\int_{\vec{k}'\omega'}^>G(\vec{k}',\omega')M_jG^T(-\vec{k}',-\omega').
\end{equation}
Overall, these contributions lead to RG equations of the form,
\begin{equation}
\frac{d\Delta_i}{d\ell}=\Delta_i+\sum_{jk}\sum_a\tilde{B}_{ijk}^{(a)}\Delta_jg_k\Phi_a(\mu,m,T).
\end{equation}

We now consider the behavior of the source terms for large $\ell$ and at $T=T_c$.  For reasons stated earlier, we will focus only on the
case where $T$ increases more rapidly than $\mu$ and $m$.  As stated before, the $\Phi_a$ functions that decrease the most slowly in this
case are $\Phi_{2,+}\approx\Phi_{2,-}\approx\frac{\Lambda}{8\pi T}$.  The equations for the $\Delta_i$ then become, after substituting the
large $\ell$ expression for the $g_i$,
\begin{equation}
\frac{d\Delta_i}{d\ell}=\Delta_i+\frac{2}{\mathcal{A}_r}\sum_j\tilde{\mathcal{B}}_{ij}\Delta_j,
\end{equation}
where
\begin{equation}
\tilde{\mathcal{B}}_{ij}=\sum_k[\tilde{B}_{ijk}^{(2,+)}+\tilde{B}_{ijk}^{(2,-)}]\rho_k.
\end{equation}
We will thus have, at most, a system of two linear equations describing a given source term.  If a given term does not couple to another, then
it will simply be given by $\Delta_i(\ell)=\Delta_i(\ell_0)e^{\eta_i\ell}$, where
\begin{equation}
\eta_i=1+\frac{2\tilde{\mathcal{B}}_{ii}}{\mathcal{A}_r}.
\end{equation}
Otherwise, we simply solve the system of equations using standard techniques.

\subsection{Free energy}
\begin{figure}
	\centering
		\includegraphics[width=\columnwidth]{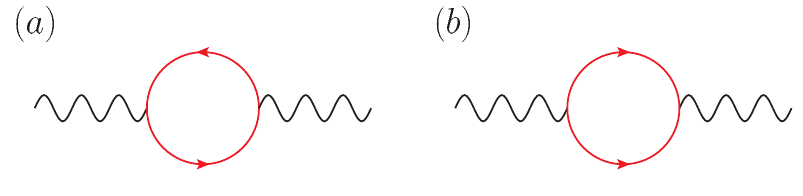}
		\caption{Diagrams representing one-loop contributions to the free energy from the (a) particle-hole source terms and (b) particle-particle
		source terms.  The wavy lines represent the source term vertices and the red lines to fast electronic modes.}
	\label{fig:FreeEnergy_OneLoop}
\end{figure}

Now that we have derived the RG equations, we turn our attention to the free energy.  More specifically, we will calculate the contribution
to the free energy per unit area from the source terms alone.  The diagrams that represent contributions from the source terms is shown
in Fig.\ \ref{fig:FreeEnergy_OneLoop}.
The first diagram, Fig.\ \ref{fig:FreeEnergy_OneLoop}a, represents the contribution from the ph source terms, and yields the following
result for the free energy per unit area from the source terms:
\begin{eqnarray}
&&\Delta f_{\Delta}(\text{ph})=-\tfrac{1}{2}\int_0^\infty d\ell\,e^{-3\ell} \cr
&&\times\int_{\vec{k}\omega}^>\sum_{ij}\Delta_i^{\text{ph}}\Delta_j^{\text{ph}}\mbox{Tr}[M_iG(\vec{k},\omega)M_jG(\vec{k},\omega)] \cr
&&=-\tfrac{1}{2}\int_0^\infty d\ell\,e^{-3\ell}\sum_{ij}\sum_a\alpha_{ij}^{(a),\text{ph}}\Delta_i^{\text{ph}}\Delta_j^{\text{ph}}\Phi_a(\mu,m,T). \nonumber \\
\end{eqnarray}
The second diagram, Fig.\ \ref{fig:FreeEnergy_OneLoop}b, represents the contribution from the pp terms, and yields
\begin{eqnarray}
&&\Delta f_{\Delta}(\text{pp})=-\tfrac{1}{4}\int_0^\infty d\ell\,e^{-3\ell} \cr
&&\times\int_{\vec{k}\omega}^>\sum_{ij}\Delta_i^{\text{pp}}(\Delta_j^{\text{pp}})^\ast\mbox{Tr}[M_iG(\vec{k},\omega)M_jG(\vec{k},\omega)]+\text{c.c.} \cr
&&=-\tfrac{1}{4}\int_0^\infty d\ell\,e^{-3\ell}\sum_{ij}\sum_a\alpha_{ij}^{(a),\text{pp}}\Delta_i^{\text{pp}}(\Delta_j^{\text{pp}})^\ast\Phi_a(\mu,m,T) \cr
&&+\text{c.c.}
\end{eqnarray}

With these results, we may now derive the susceptibilities and thus determine which of them diverge for a given fixed ray.  More
specifically, we will determine their behavior for $T$ close to, and just above, $T_c$.  We start by revisiting the equations for
the $g_i$.  In this case, we may still use the large $\ell$ approximations for the $\Phi_a$, but now we assume that $g_i$ tends
to a very large, but finite, value as $\ell\to\infty$.  If we solve the equation for the $g_r$ that we divide by to obtain the
ratios $\rho_i$ in this case, we get
\begin{equation}
\frac{1}{g_r(\ell,T)}=\frac{e^{\ell-\ell_0}}{g_r(\ell_0,T)}+\frac{\mathcal{A}_r}{16\pi T}(e^{-2\ell}-e^{-2\ell_0})e^\ell,
\end{equation}
where $\ell_0\gg 1$.  We now make use of the fact that, at $T=T_c$,
\begin{equation}
\frac{1}{g_r(\ell_0,T_c)}=\frac{\mathcal{A}_r}{16\pi T_c}e^{-\ell_0},
\end{equation}
to rewrite the above as
\begin{eqnarray}
\frac{1}{g_r(\ell,T)}&=&e^{\ell-\ell_0}\left [\frac{1}{g_r(\ell_0,T)}-\frac{1}{g_r(\ell_0,T_c)}\right ] \cr
&+&\frac{\mathcal{A}_r}{16\pi}e^{\ell-2\ell_0}\left (\frac{1}{T_c}-\frac{1}{T}\right )+\frac{\mathcal{A}_r}{16\pi T}e^{-\ell}.
\end{eqnarray}
We may now use the fact that, to first order in $T-T_c$,
\begin{equation}
\frac{1}{g_r(\ell_0,T)}\approx\frac{1}{g_r(\ell_0,T_c)}+\left.\left [\frac{\partial}{\partial T}\frac{1}{g_r(\ell_0,T)}\right ]\right |_{T=T_c}(T-T_c)
\end{equation}
to obtain
\begin{equation}
\frac{1}{g_r(\ell,T)}=c_r(T-T_c)e^\ell+\frac{\mathcal{A}_r}{16\pi T}e^{-\ell},
\end{equation}
where $c_r$ is a constant.

We now perform a similar analysis of the equations for the source terms.  Doing so, we find that, for a $\Delta_i$ that is
not coupled to any other $\Delta_j$,
\begin{equation}
\Delta_i(\ell,T)=\Delta_i(\ell_0,T)e^{\eta_i(\ell-\ell_0)}\left [\frac{C_r(T-T_c)e^{2\ell_0}+1}{C_r(T-T_c)e^{2\ell}+1}\right ]^{(\eta_i-1)/2},
\end{equation}
where $C_r$ is a constant proportional to the constant $c_r$ appearing in the equation for $g_r(\ell,T)$.

If we now substitute these results into the free energy and determine the contribution from the integral on $\ell$ for $\ell>\ell_0$
(which is where we expect the divergence to come from), we find that it goes as $(T-T_c)^{2-\eta_i}$, and thus the susceptibility,
\begin{equation}
\chi_{ii}^{\text{ph}}(T)=-\frac{\partial^2f}{\partial[\Delta_i^{\text{ph}}(\ell=0,T)]^2}
\end{equation}
or
\begin{equation}
\chi_{ii}^{\text{pp}}(T)=-\frac{\partial^2f}{\partial\Delta_i^{\text{pp}}(\ell=0,T)\partial[\Delta_i^{\text{pp}}(\ell=0,T)]^\ast},
\end{equation}
will diverge with the same exponent provided that $\eta_i>2$.  We note that $\Delta_i(\ell=\ell_0)\propto\Delta_i(\ell=0)$ simply due to
the linearity of the equations giving $\Delta_i$.  Therefore, if the condition that $\eta_i>2$, or
\begin{equation}
\frac{2\mathcal{B}_{ii}}{A_r}>1,
\end{equation}
is satisfied, then the corresponding susceptibility exponent diverges, and we claim that the system is unstable
to the associated order.

A similar analysis for the case of coupled source terms shows that, provided that the system of RG equations yields
at least one solution $e^{\eta_i\ell}$ that satisfies the condition that $\eta_i>2$, then we have an instability
towards the corresponding order.  We note that coupling of source terms only occurs if they correspond to the same
representation of the symmetry group of the system.

With these results, we are now ready to determine the leading instabilities of the system that the various fixed
rays correspond to.  Unlike in the case of bilayer graphene, considered in Refs.\ \onlinecite{ThrockmortonPRB2014} and
\onlinecite{CvetkovicPRB2012}, we do not find any one-parameter or more families of fixed rays; all of the fixed
rays are isolated.  However, we find a very large number (thousands) of solutions.  A number of these rays correspond
to multiple instabilities simultaneously present.  The fact that we only obtain isolated fixed rays, rather than any
multiparameter families, indicates that the system is unstable to perturbations in the initial conditions, causing
the system to converge to a different fixed ray.  These two facts are not surprising, given the diversity of symmetry-breaking
states found so far in the related, but not identical, twisted bilayer graphene system.  We also find one of two
outcomes for the mass $m$ and the chemical potential $\mu$---either the mass diverges to infinity and the chemical
potential goes to zero, or vice versa.  These two outcomes would result from starting with the chemical potential
inside or outside the gap, respectively.

\section{Conclusion}
We have investigated the possibility of instabilities to interaction-induced symmetry-breaking orders in a honeycomb lattice away from
half-filling, i.e., the chemical potential $\mu\neq 0$, subject to a periodic potential.  For simplicity, we assumed
that the periodic potential preserved a $D_3$ point group symmetry, along with translational symmetry (though reduced
from that of the unmodified lattice), time reversal, and spin $SU(2)$.  This allows for a mass gap $m$ to be present.
We employ a finite-temperature Wilson-Fisher momentum shell RG procedure in this work.  We derived the RG equations
for the four-fermion coupling constants $g_i$, the chemical potential $\mu$, the mass $m$, and the temperature $T$.
Also for simplicity, we focused on the case in which $T$ increases more quickly at large RG scaling parameter $\ell$
than $\mu$ or $m$.  We then showed that, at some ``critical temperature'' $T_c$, the coupling constants diverge exponentially,
but that ratios thereof remained finite.  We finally showed how these ratios could be used to determine what symmetry-breaking
orders the system would be unstable to.  We found that there were thousands of isolated fixed rays, a contrast to
similar studies of bilayer graphene\cite{ThrockmortonPRB2014,CvetkovicPRB2012}, where a two-parameter family of fixed
rays was found in addition to only a few isolated rays.  In some cases, these rays corresponded to instabilities toward
several different orders, which is not surprising given the diverse orders detected in experiments on the related, though
not entirely identical, twisted bilayer graphene.

Our detailed multi-parameter RG analysis within a simple minimal lattice model points to the real possibility that moir\'e
superlattices (e.g. twisted bilayer graphene near the magic angle), may manifest "unstable symmetry-breaking" where the
symmetry-breaking phases are intrinsically fragile, and the physics depends sensitively on all the details and initial
conditions, even excluding the realistic complications of disorder, strain, phonon, substrate, etc.  Our work is consistent
with there being considerable sample dependence in the observed phenomenology of various correlated exotic phases in tBLG.
Such a generic fragile ``unstable symmetry breaking'' scenario leads us to conclude that it is likely that experimental
development would lead to the observation of many exotic ground states in the system.

Similar considerations in these previous works on bilayer graphene concerning the exponents apply here as well; we
expect that our procedure captures the basic qualitative behavior of the susceptibilities (i.e., whether or not they
diverge), even if the exact exponents are not quite correct.  We also note that many of these fixed rays correspond
to multiple instabilities.  Our method does not give further information other than the possibility of these orders
emerging.  Other methods are required to determine which of these orders actually emerges.

\acknowledgments
This work was funded by the Laboratory for Physical Sciences.

\appendix

\section{Exact non-interacting Hamiltonian} \label{App:ExactH}
\begin{figure}
	\centering
		\includegraphics[width=0.49\columnwidth]{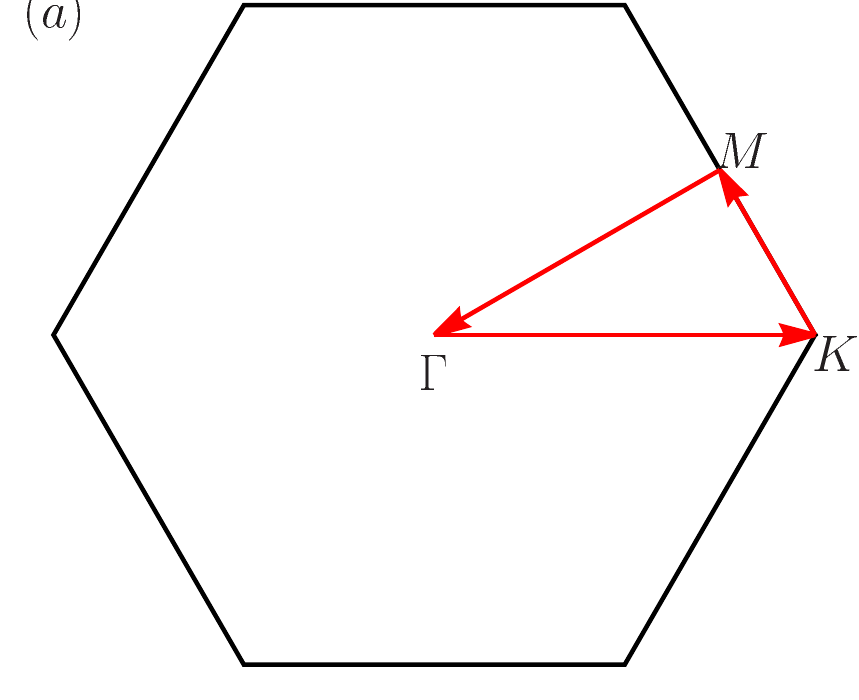}
		\includegraphics[width=0.49\columnwidth]{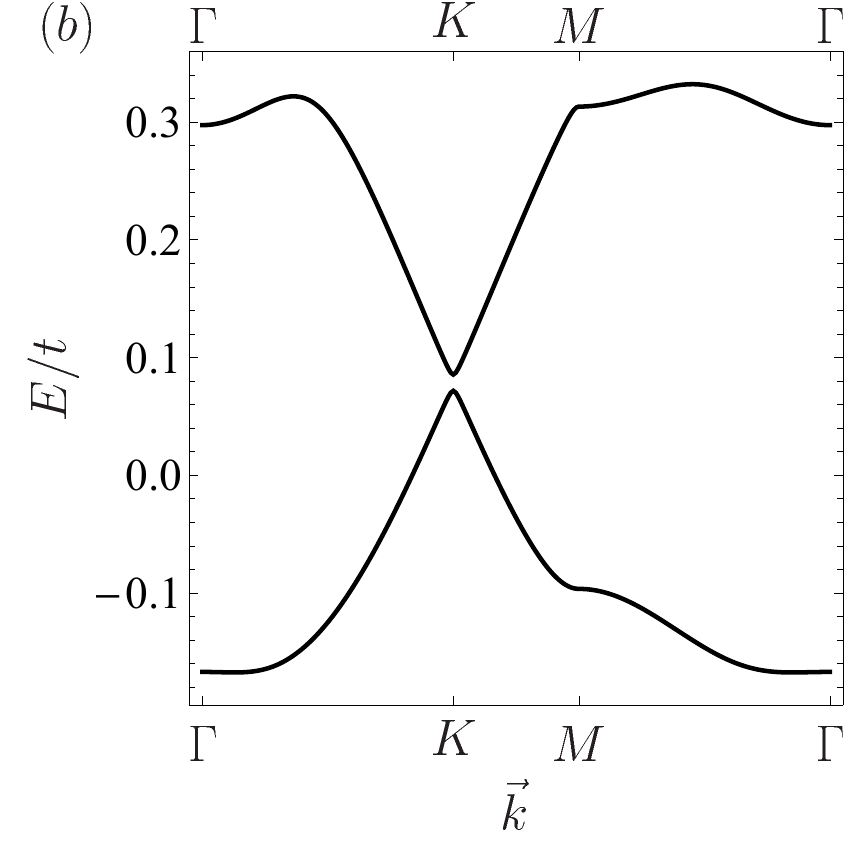}
	\caption{(a) Illustration of the Brillouin zone for the full Hamiltonian, Eq.\ \eqref{Eq:HCLattice_PP_Exact}.  The
	path along which we sweep the crystal momentum $\vec{k}$ in the plot to the right is shown in red.  (b) Plot of
	the middle two bands obtained from the Hamiltonian, Eq.\ \eqref{Eq:HCLattice_PP_Exact}, 	as a function of crystal
	momentum $\vec{k}$.}
	\label{fig:Graphene_PPot_BandStruct}
\end{figure}

We show here that the exact band structure resulting from a tight-binding honeycomb lattice Hamiltonian with an
external periodic potential produces the features that we incorporate into our low-energy effective theory.  The
exact Hamiltonian is given by
\begin{equation}
H=-t\sum_{\vec{r},\sigma}[c_\sigma^\dag(\vec{r})c_\sigma(\vec{r}+\vec{\delta})+\text{h.c.}]+\sum_{\vec{r},\sigma}V(\vec{r})n_\sigma(\vec{r}), \label{Eq:HCLattice_PP_Exact}
\end{equation}
where $n_\sigma(\vec{r})=c_\sigma^\dag(\vec{r})c_\sigma(\vec{r})$, $V(\vec{r})$ is a periodic potential that we
assume has a periodicity commensurate with the lattice, $\vec{r}$ runs over the underlying triangular Bravais lattice
for the honeycomb lattice, $\sigma$ is the spin, and $\vec{\delta}=a\hat{\vec{y}}$.  If $\vec{a}_1$ and $\vec{a}_2$
are the primitive vectors of the honeycomb lattice's Bravais lattice, then we assume that the corresponding primitive
vectors of the periodic potential are $n\vec{a}_1$ and $n\vec{a}_2$.  In general, even the constraints that we assume
on the periodic potential can eliminate all point-group symmetries, leaving only translational, time reversal, and
spin $SU(2)$ symmetries.  However, certain choices of potential will preserve up to a $D_3$ point-group symmetry, the
situation that we focus on in our work.  One such potential is
\begin{equation}
V(\vec{r})=V_0[\cos(\vec{Q}_1\cdot\vec{r})+\cos(\vec{Q}_2\cdot\vec{r})+\cos(\vec{Q}_3\cdot\vec{r})],
\end{equation}
where $\vec{Q}_1=\frac{2\pi}{n\sqrt{3}a}\hat{\vec{x}}-\frac{2\pi}{3na}\hat{\vec{y}}$, $\vec{Q}_2=\frac{4\pi}{3na}\hat{\vec{y}}$,
and $\vec{Q}_3=-\vec{Q}_1-\vec{Q}_2=-\frac{2\pi}{n\sqrt{3}a}\hat{\vec{x}}-\frac{2\pi}{3na}\hat{\vec{y}}$.  To show
the existence of Dirac cones in this model, we numerically diagonalize the Hamiltonian and show the middle two bands,
which we do for $n=5$ and $V_0=t$ in Fig.\ \ref{fig:Graphene_PPot_BandStruct}.  We see that there is a gapped Dirac
cone at crystal momentum $\vec{K}=\frac{4\pi}{3n\sqrt{3}}\hat{\vec{x}}$.

\section{Integrals occurring in the RG equations for the four-fermion coupling constants} \label{App:Ints_gRGEqus}
We now present the sums and integrals that occur in evaluating the integral and sum in Eq.\ \eqref{Eq:2GFIntSum}.
Doing so, we obtain
\begin{eqnarray}
&&\int_{\vec{k}\omega}^>G_0(\vec{k},\omega)\otimes G_0(\vec{k},\omega) \cr
&&=\Phi_{2,+}1\otimes 1+I_{1,z,+}(1\otimes 1\sigma_z1+1\sigma_z1\otimes 1) \cr
&&+\Phi_{0,xy,+}(\tau_z\sigma_x1\otimes \tau_z\sigma_x1+1\sigma_y1\otimes 1\sigma_y1) \cr
&&+\Phi_{0,z,+}1\sigma_z1\otimes 1\sigma_z1 \label{Eq:2GF_Plus}
\end{eqnarray}
and
\begin{eqnarray}
&&\int_{\vec{k}\omega}^>G_0(\vec{k},\omega)\otimes G_0(-\vec{k},-\omega) \cr
&&=\Phi_{2,-}1\otimes 1+I_{1,z,-+}1\otimes 1\sigma_z1+I_{1,z,--}1\sigma_z1\otimes 1 \cr
&&+\Phi_{0,xy,-}(\tau_z\sigma_x1\otimes \tau_z\sigma_x1+1\sigma_y1\otimes 1\sigma_y1) \cr
&&+\Phi_{0,z,-}1\sigma_z1\otimes 1\sigma_z1, \label{Eq:2GF_Minus}
\end{eqnarray}
where the functions $\Phi_a$, which also occur in \eqref{Eq:gRGEqus}, are given by 
\begin{widetext}
\begin{eqnarray}
\Phi_{0,xy,+}&=&\int_{\vec{k}\omega}^>\frac{v_F^2k_x^2}{\{(-\omega+i\mu)^2+[E(\vec{k})]^2\}^2}=\int_{\vec{k}\omega}^>\frac{v_F^2k_y^2}{\{(-\omega+i\mu)^2+[E(\vec{k})]^2\}^2} \cr
&=&\frac{v_F^2\Lambda}{32\pi E_\Lambda^3}\left [\tanh\left (\frac{E_\Lambda-\mu}{2T}\right )+\tanh\left (\frac{E_\Lambda+\mu}{2T}\right )\right ]-\frac{v_F^2\Lambda}{64\pi E_\Lambda^2T}\left [\frac{1}{\cosh^2\left (\frac{E_\Lambda-\mu}{2T}\right )}+\frac{1}{\cosh^2\left (\frac{E_\Lambda+\mu}{2T}\right )}\right ], \nonumber \\ \\
\Phi_{0,z,+}&=&\int_{\vec{k}\omega}^>\frac{m^2}{\{(-\omega+i\mu)^2+[E(\vec{k})]^2\}^2} \cr
&=&\frac{m^2\Lambda}{16\pi E_\Lambda^3}\left [\tanh\left (\frac{E_\Lambda-\mu}{2T}\right )+\tanh\left (\frac{E_\Lambda+\mu}{2T}\right )\right ]-\frac{m^2\Lambda}{32\pi E_\Lambda^2T}\left [\frac{1}{\cosh^2\left (\frac{E_\Lambda-\mu}{2T}\right )}+\frac{1}{\cosh^2\left (\frac{E_\Lambda+\mu}{2T}\right )}\right ], \nonumber \\ \\
\Phi_{1,z,+}&=&\int_{\vec{k}\omega}^>\frac{m(i\omega+\mu)}{\{(-\omega+i\mu)^2+[E(\vec{k})]^2\}^2}=\frac{m\Lambda}{32\pi E_\Lambda T}\left [\frac{1}{\cosh^2\left (\frac{E_\Lambda-\mu}{2T}\right )}-\frac{1}{\cosh^2\left (\frac{E_\Lambda+\mu}{2T}\right )}\right ], \\
\Phi_{2,+}&=&\int_{\vec{k}\omega}^>\frac{(i\omega+\mu)^2}{\{(-\omega+i\mu)^2+[E(\vec{k})]^2\}^2} \cr
&=&\frac{\Lambda}{16\pi E_\Lambda}\left [\tanh\left (\frac{E_\Lambda-\mu}{2T}\right )+\tanh\left (\frac{E_\Lambda+\mu}{2T}\right )\right ]+\frac{\Lambda}{32\pi T}\left [\frac{1}{\cosh^2\left (\frac{E_\Lambda-\mu}{2T}\right )}+\frac{1}{\cosh^2\left (\frac{E_\Lambda+\mu}{2T}\right )}\right ], \nonumber \\ \\
\Phi_{0,xy,-}&=&\int_{\vec{k}\omega}^>\frac{v_F^2k_x^2}{\{(-\omega+i\mu)^2+[E(\vec{k})]^2\}\{(\omega+i\mu)^2+[E(\vec{k})]^2\}}=\int_{\vec{k}\omega}^>\frac{v_F^2k_y^2}{\{(-\omega+i\mu)^2+[E(\vec{k})]^2\}\{(\omega+i\mu)^2+[E(\vec{k})]^2\}} \cr
&=&-\frac{\Lambda}{32\pi\mu E_\Lambda}\left [\frac{1}{E_\Lambda-\mu}\tanh\left (\frac{E_\Lambda-\mu}{2T}\right )-\frac{1}{E_\Lambda+\mu}\tanh\left (\frac{E_\Lambda+\mu}{2T}\right )\right ], \\
\Phi_{0,z,-}&=&\int_{\vec{k}\omega}^>\frac{m^2}{\{(-\omega+i\mu)^2+[E(\vec{k})]^2\}\{(\omega+i\mu)^2+[E(\vec{k})]^2\}} \cr
&=&\frac{m^2\Lambda}{16\pi\mu E_\Lambda}\left [\frac{1}{E_\Lambda-\mu}\tanh\left (\frac{E_\Lambda-\mu}{2T}\right )-\frac{1}{E_\Lambda+\mu}\tanh\left (\frac{E_\Lambda+\mu}{2T}\right )\right ], \\
\Phi_{1,z,-+}&=&\int_{\vec{k}\omega}^>\frac{m(i\omega+\mu)}{\{(-\omega+i\mu)^2+[E(\vec{k})]^2\}\{(\omega+i\mu)^2+[E(\vec{k})]^2\}} \cr
&=&-\frac{m\Lambda}{16\pi\mu E_\Lambda}\left [\frac{1}{E_\Lambda-\mu}\tanh\left (\frac{E_\Lambda-\mu}{2T}\right )-\frac{1}{E_\Lambda+\mu}\tanh\left (\frac{E_\Lambda+\mu}{2T}\right )\right ], \\
\Phi_{1,z,--}&=&\int_{\vec{k}\omega}^>\frac{m(-i\omega+\mu)}{\{(-\omega+i\mu)^2+[E(\vec{k})]^2\}\{(\omega+i\mu)^2+[E(\vec{k})]^2\}}=-\Phi_{1,z,-+}, \\
\Phi_{2,-}&=&\int_{\vec{k}\omega}^>\frac{(-\omega+i\mu)(\omega+i\mu)}{\{(-\omega+i\mu)^2+[E(\vec{k})]^2\}\{(\omega+i\mu)^2+[E(\vec{k})]^2\}} \cr
&=&\frac{\Lambda}{16\pi\mu}\left [\frac{E_\Lambda-2\mu}{E_\Lambda-\mu}\tanh\left (\frac{E_\Lambda-\mu}{2T}\right )-\frac{E_\Lambda+2\mu}{E_\Lambda+\mu}\tanh\left (\frac{E_\Lambda+\mu}{2T}\right )\right ]
\end{eqnarray}
\end{widetext}

\section{Contributions to the RG equations for the four-fermion coupling constants} \label{App:Coeffs_gRGEqus}
We now consider each diagram in turn, starting with the ``bubble'' diagram, Fig.\ \ref{fig:FourFermion_OneLoop}a.
Evaluating this contribution, we obtain
\begin{eqnarray}
&&\Delta S_2(\text{bubble}) \cr
&&=-\tfrac{1}{8}\sum_{ij}g_ig_j\int_{1234}\mbox{Tr}\left\{\int_{\vec{k}\omega}^>G(\vec{k},\omega)S_{i,2}G(\vec{k},\omega)S_{j,1}\right\} \cr
&&\times[\psi^\dag(1)S_{i,1}\psi(2)][\psi^\dag(3)S_{j,2}\psi(4)].
\end{eqnarray}

Next, we consider the ``side interaction'' diagrams, Figs.\ \ref{fig:FourFermion_OneLoop}b and \ref{fig:FourFermion_OneLoop}c.
These together give us the following contribution:
\begin{eqnarray}
&&\Delta S_2(\text{side int.}) \cr
&&=\tfrac{1}{8}\sum_{ij}g_ig_j\int_{1234}[\psi^\dag(1)S_{i,1}\psi(2)][\psi^\dag(3)M_{i,1}\psi(4)] \cr
&&+\tfrac{1}{8}\sum_{ij}g_ig_j\int_{1234}[\psi^\dag(1)S_{i,2}\psi(2)][\psi^\dag(3)M_{i,2}\psi(4)] \cr
&&+\tfrac{1}{8}\sum_{ij}g_ig_j\int_{1234}[\psi^\dag(1)S_{j,1}\psi(2)][\psi^\dag(3)M_{j,1}\psi(4)] \cr
&&+\tfrac{1}{8}\sum_{ij}g_ig_j\int_{1234}[\psi^\dag(1)S_{j,2}\psi(2)][\psi^\dag(3)M_{j,2}\psi(4)], \nonumber \\
\end{eqnarray}
where
\begin{eqnarray}
M_{i,1}&=&\int_{\vec{k}\omega}^>[S_{j,1}G(\vec{k},\omega)S_{i,2}G(\vec{k},\omega)S_{j,2} \cr
&+&S_{j,2}G(\vec{k},\omega)S_{i,2}G(\vec{k},\omega)S_{j,1}], \\
M_{i,2}&=&\int_{\vec{k}\omega}^>[S_{j,1}G(\vec{k},\omega)S_{i,1}G(\vec{k},\omega)S_{j,2} \cr
&+&S_{j,2}G(\vec{k},\omega)S_{i,1}G(\vec{k},\omega)S_{j,1}], \\
M_{j,1}&=&\int_{\vec{k}\omega}^>[S_{i,1}G(\vec{k},\omega)S_{j,2}G(\vec{k},\omega)S_{i,2} \cr
&+&S_{i,2}G(\vec{k},\omega)S_{j,2}G(\vec{k},\omega)S_{i,1}], \\
M_{j,2}&=&\int_{\vec{k}\omega}^>[S_{i,1}G(\vec{k},\omega)S_{j,1}G(\vec{k},\omega)S_{i,2} \cr
&+&S_{i,2}G(\vec{k},\omega)S_{j,1}G(\vec{k},\omega)S_{i,1}].
\end{eqnarray}

Next, we consider the first ``ladder'' diagram, Fig.\ \ref{fig:FourFermion_OneLoop}d.  This diagram yields the following contribution:
\begin{eqnarray}
&&\Delta S_2(\text{ladder 1}) \cr
&&=\tfrac{1}{8}\sum_{ij}g_ig_j\int_{1234}\int_{\vec{k}\omega}^>[\psi^\dag(1)S_{j,1}G(\vec{k},\omega)S_{i,1}\psi(2)] \cr
&&\times[\psi^\dag(3)S_{j,2}G(-\vec{k},-\omega)S_{i,2}\psi(4)] \cr
&&+\tfrac{1}{8}\sum_{ij}g_ig_j\int_{1234}\int_{\vec{k}\omega}^>[\psi^\dag(1)S_{j,1}G(\vec{k},\omega)S_{i,2}\psi(2)] \cr
&&\times[\psi^\dag(3)S_{j,2}G(-\vec{k},-\omega)S_{i,1}\psi(4)] \cr
&&+\tfrac{1}{8}\sum_{ij}g_ig_j\int_{1234}\int_{\vec{k}\omega}^>[\psi^\dag(1)S_{i,1}G(\vec{k},\omega)S_{j,1}\psi(2)] \cr
&&\times[\psi^\dag(3)S_{i,2}G(-\vec{k},-\omega)S_{j,2}\psi(4)] \cr
&&+\tfrac{1}{8}\sum_{ij}g_ig_j\int_{1234}\int_{\vec{k}\omega}^>[\psi^\dag(1)S_{i,1}G(\vec{k},\omega)S_{j,2}\psi(2)] \cr
&&\times[\psi^\dag(3)S_{i,2}G(-\vec{k},-\omega)S_{j,1}\psi(4)].
\end{eqnarray}

Finally, the second ``ladder'' diagram, Fig.\ \ref{fig:FourFermion_OneLoop}e, yields
\begin{eqnarray}
&&\Delta S_2(\text{ladder 2}) \cr
&&=\tfrac{1}{8}\sum_{ij}g_ig_j\int_{1234}\int_{\vec{k}\omega}^>[\psi^\dag(1)S_{j,1}G(\vec{k},\omega)S_{i,1}\psi(2)] \cr
&&\times[\psi^\dag(3)S_{i,2}G(\vec{k},\omega)S_{j,2}\psi(4)] \cr
&&+\tfrac{1}{8}\sum_{ij}g_ig_j\int_{1234}\int_{\vec{k}\omega}^>[\psi^\dag(1)S_{j,1}G(\vec{k},\omega)S_{i,2}\psi(2)] \cr
&&\times[\psi^\dag(3)S_{i,2}G(\vec{k},\omega)S_{j,1}\psi(4)] \cr
&&+\tfrac{1}{8}\sum_{ij}g_ig_j\int_{1234}\int_{\vec{k}\omega}^>[\psi^\dag(1)S_{i,1}G(\vec{k},\omega)S_{j,1}\psi(2)] \cr
&&\times[\psi^\dag(3)S_{j,2}G(\vec{k},\omega)S_{i,2}\psi(4)] \cr
&&+\tfrac{1}{8}\sum_{ij}g_ig_j\int_{1234}\int_{\vec{k}\omega}^>[\psi^\dag(1)S_{i,1}G(\vec{k},\omega)S_{j,2}\psi(2)] \cr
&&\times[\psi^\dag(3)S_{j,2}G(\vec{k},\omega)S_{i,1}\psi(4)].
\end{eqnarray}

We may then determine which terms these contributions renormalize with the aid of Eqs.\ \eqref{Eq:2GF_Plus} and \eqref{Eq:2GF_Minus}
and the standard $SU(8)$ identity,
\begin{equation}
A=\tfrac{1}{8}\sum_n\mbox{Tr}(AM_n)M_n,
\end{equation}
where $M_n$ runs over all matrices of the form, $\tau_i\sigma_js_k$.

\end{document}